\documentclass[10pt,journal]{IEEEtran}
\IEEEoverridecommandlockouts

\usepackage{graphicx,subfigure}
\usepackage{amsmath,amsthm,amsfonts,amssymb}
\usepackage{cite}
\usepackage{bm}
\usepackage{bbm}
\usepackage{url}
\usepackage{array}
\usepackage{color}
\usepackage{multirow}
\usepackage{booktabs,multirow}
\usepackage[table,xcdraw]{xcolor}
\usepackage{enumerate}
\usepackage{enumitem}
\usepackage{makecell}

\usepackage[english]{babel}
\usepackage{algorithm}
\usepackage[noend]{algpseudocode}

\theoremstyle{plain}

\usepackage{mathtools}
\DeclarePairedDelimiter{\ceil}{\lceil}{\rceil}

\usepackage[english]{babel}
\usepackage{algorithm}
\usepackage[noend]{algpseudocode}

\allowdisplaybreaks[4]

\begin{document}
% \title{Deep Learning-Aided Wireless Communications with Discrete-time Analog Transmission: A PAPR Perspective}
\title{Semantic Communications with Discrete-time Analog Transmission: A PAPR Perspective}

\author{
Yulin~Shao,~\IEEEmembership{Member,~IEEE},
Deniz~G\"und\"uz,~\IEEEmembership{Fellow,~IEEE}
\thanks{Y. Shao and D. G\"und\"uz are with the Department of Electrical and Electronic Engineering, Imperial College London, London SW7 2AZ, U.K. (e-mail: \{y.shao,d.gunduz\}@imperial.ac.uk).
}
}

\maketitle

\begin{abstract}
Recent progress in deep learning (DL)-based joint source-channel coding (DeepJSCC) has led to a new paradigm of semantic communications.
Two salient features of DeepJSCC-based semantic communications are the exploitation of semantic-aware features directly from the source signal, and the discrete-time analog transmission (DTAT) of these features.
Compared with traditional digital communications, semantic communications with DeepJSCC provide superior reconstruction performance at the receiver and graceful degradation with diminishing channel quality, but also exhibit a large peak-to-average power ratio (PAPR) in the transmitted signal.
An open question has been whether the gains of DeepJSCC come from the additional freedom brought by the high-PAPR continuous-amplitude signal.
In this paper, we address this question by exploring three PAPR reduction techniques in the application of image transmission. We confirm that the superior image reconstruction performance of DeepJSCC-based semantic communications can be retained while the transmitted PAPR is suppressed to an acceptable level. This observation is an important step towards the implementation of DeepJSCC in practical semantic communication systems.
\end{abstract}


\section{Introduction}\label{sec:intro}
There has been a growing interest in developing new semantic-aware communication systems \cite{Deniz2022,semanticTheory,JSCC2019,xie2021} via data-driven approaches such as deep learning (DL). Compared with legacy digital communications, DL-enabled semantic communications leverage deep joint source-channel coding (DeepJSCC) \cite{JSCC2019} and extract semantic-aware and goal-oriented information directly from the source, yielding better source reconstruction.

An important ingredient of DL-enabled wireless communication systems is discrete-time analog transmission (DTAT) \cite{xie2021,JSCC2019,Haotian,DeepJSCCq,yang2022ofdm,AttentionCode,KO2021,FLOAC}. Specifically, deep neural network (DNN)-based encoder and decoder are capable of exploiting discrete-time continuous amplitude signals, yielding more freedom than discrete constellations. Increasingly more evidence reveals that DTAT contributes a large part to the excellent performance of DL-enabled communication systems. For example, in DL-aided channel coding (where the source is a stream of bits), the gains of neural channel codes almost vanish when the coded symbols are limited to binary inputs, e.g., BPSK \cite{KO2021}. Also, in semantic image transmission, the image reconstruction performance deteriorates significantly when the JSCC-coded symbols are limited to a set of discrete constellations \cite{DeepJSCCq} -- high-quality reconstruction is possible only when the constellation size is extremely large, resembling continuous amplitude signals.

The practical use of DTAT, however, faces an important challenge due to the peak-to-average power ratio (PAPR) \cite{PAPRsurvey}, especially when used in conjunction with the prevailing orthogonal frequency division multiplexing (OFDM) transceiver, which is now adopted in most IEEE standards. PAPR stems from the saturation effect of the power amplifier: a low PAPR of the transmitted signal is desired as the power amplifier can operate more efficiently and the coverage of the transmission is larger. Said in another way, for a given communication range, low-PAPR signals save the transmission power. Thus, PAPR is more critical in the uplink transmission of a mobile communication system since the battery power of mobile users is limited.

{\it Related work:}
In traditional digital communications, the OFDM signal exhibits a large PAPR since independent quadrature amplitude modulation (QAM)-modulated waveforms are linearly combined. To address this problem, various methods, such as clipping, coding, scrambling, and linear precoding, have been proposed in the literature. We refer readers to \cite{PAPRsurvey} for a detailed review.  In addition to traditional schemes, DL-based approaches to reducing the PAPR of OFDM signals are proposed in \cite{PAPRloss1,PAPRloss3,PAPRloss4}. Specifically, these works focus on the transmission of independent and identically distributed (i.i.d.) constellations. The OFDM system is modeled and trained in an end-to-end fashion, and the main idea is to incorporate the PAPR into the loss function in addition to the original decoding loss, e.g., bit error rate (BER). In so doing, the DL-based transceiver can learn to generate low-PAPR signals as training progresses.

In DL-aided semantic communications, the coded symbols have continuous amplitude. The PAPR of the transmitted signal after OFDM modulation is even more severe than that in digital communications because both DTAT and OFDM modulation contribute to the high PAPR. Prior works on semantic communications emphasize the decoding or reconstruction performance \cite{xie2021,JSCC2019,Haotian,DeepJSCCq}, while the PAPR performance has been mostly ignored so far. One exception is \cite{yang2022ofdm}, wherein the authors focused on semantic image transmission with OFDM and investigated the impact of clipping on the image reconstruction performance of the proposed DNN architecture.

It is worth noting that all the works mentioned above evaluate the PAPR using baseband complex symbols directly after OFDM modulation, which does not accurately reflect the PAPR properties of passband real signals after pulse shaping and frequency upshifting. 

{\it Contributions:}
In this paper, we study the PAPR performance of DL-aided semantic communications considering the uplink transmission from mobile users to the base station in cellular networks. The cellular network is operated with orthogonal frequency-division multiple access (OFDMA), where the radio channel is divided into multiple subchannels/subcarriers and each user is allocated with a subset of subcarriers. 
We develop a passband transceiver for the OFDMA system and – unlike prior works – evaluate the PAPR at the passband. Note that our passband transceiver\footnote{Our code is available at \url{https://github.com/lynshao/SemanticPAPR}.} can be readily incorporated into any other DL-aided communication systems to evaluate the PAPR performance.

We show that there is a trade-off between achieving low-PAPR transmission and high-quality reconstruction. Focusing on the application of semantic image transmission, we investigate three PAPR reduction techniques, i.e., linearly precoded OFDMA, clipping, and PAPR loss, and characterize the trade-off between PAPR and peak signal-to-noise ratio (PSNR) of the reconstructed image by means of the relative operating characteristic (ROC) curve. 
Simulation results show that, among the three techniques, clipping strikes the best trade-off between PAPR and PSNR, when we implement a differentiable clipping operation and incorporate clipping into the training process.
DeepJSCC can then adapt to the inter-carrier interference (ICI) caused by clipping and achieve high PSNR even when the PAPR of the continuous-amplitude signal is suppressed at an acceptable level.
In other words, the superior gains of DeepJSCC with DTAT do not rely on the additional freedom brought by the high PAPR.

\section{System Model}\label{sec:II}
\begin{figure}[t]
  \centering
  \includegraphics[width=1\linewidth]{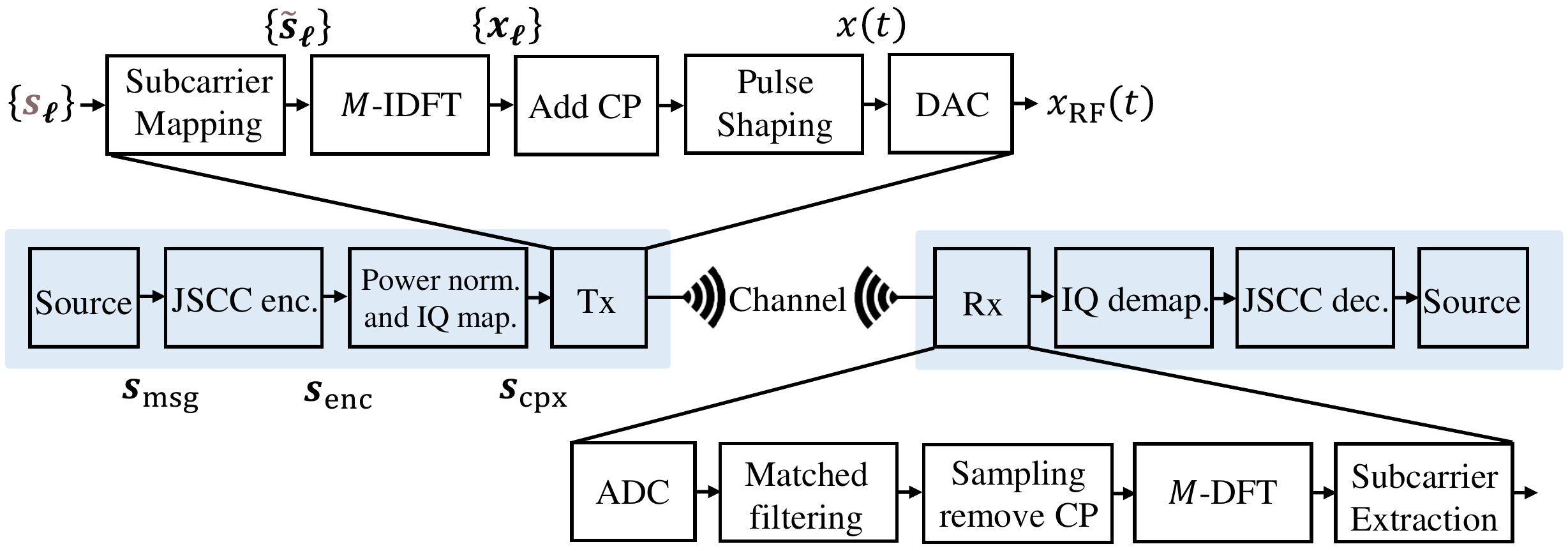}\\
  \caption{DL-enabled semantic communications with a passband transceiver.}
\label{fig:model}
\end{figure}

We consider the uplink transmission from a mobile user to the base station in an OFDMA cellular network.
The radio channel is divided into $M$ subcarriers indexed by $\mathcal{M}=\{1,2,3,...,M\}$ and the mobile user is allocated with $N$ ($N\leq M$) subcarriers indexed by $\{k_n:k_n\in \mathcal{M},n=1,2,...,N\}$.
The goal is to deliver a source message -- which can be a bitstream, an image, a text, a video, etc. -- through the physical wireless channel. The DL-enabled semantic communication system is illustrated in Fig.~\ref{fig:model}.

At the transmitter, we denote the source message by a vector $\bm{s}_{\text{msg}}$ of length $L_s$; the elements of $\bm{s}_{\text{msg}}$ can be continuous or discrete. After DeepJSCC encoding, the source message is transformed to a vector of real coded symbols $\bm{s}_{\text{enc}}\in\mathcal{R}^{L_e\times 1}$. 
To align with prior works, we define the system bandwidth ratio as $R\triangleq L_e/2 L_s$. Note that when the source message is a sequence of i.i.d. bits, the DeepJSCC encoder reduces to a deep channel encoder.

The encoded symbols are subject to an average power constraint $P$. Thus, we normalize $\bm{s}_{\text{enc}}$ such that the average power of the coded symbols generated from a source message is $P$, yielding
% \footnote{In the literature, a less stringent power constraint is $\mathbb{E}_{\bm{s}_{\text{msg}}}\left[\|\bm{s}_{\text{norm}}\|^2_2\right]\leq P$ \cite{AttentionCode,GBAF}. That is, the average power of the coded symbols across different source messages $\bm{s}_{\text{msg}}$ is less than $P$.}
\begin{equation}
\bm{s}_{\text{norm}}=\sqrt{P}\frac{\bm{s}_{\text{enc}}-\mathbb{E}[\bm{s}_{\text{enc}}]}{\|\bm{s}_{\text{enc}}-\mathbb{E}[\bm{s}_{\text{enc}}]\|_2},
\end{equation}
where $\mathbb{E}[\bm{s}_{\text{enc}}]$ is the mean of $\bm{s}_{\text{enc}}$ and $\|*\|_2$ denotes the L2 norm of a vector. Given the power normalized real vector $\bm{s}_{\text{norm}}\in\mathcal{R}^{L_e\times 1}$, we construct a complex vector $\bm{s}_{\text{cpx}}\in\mathcal{C}^{L_e/2\times 1}$, where the real and imaginary components of $\bm{s}_{\text{cpx}}$ are the odd and even elements of $\bm{s}_{\text{norm}}$.

Since the transmitter is allocated with $N$ subcarriers, we partition the complex vector $\bm{s}_{\text{cpx}}$ into $L=\ceil*{L_e/2N}$ blocks, denoted by $\{\bm{s_{\ell}}:\ell=0,1,2,...,L-1\}$, and each block $\bm{s_\ell}$ consists of $N$ complex symbols. For each block, we map $\bm{s_\ell}\in\mathcal{C}^{N\times 1}$ onto the $N$ allocated subcarriers $\{k_1,k_2,k_3,...,k_N\}$ and obtain an OFDM symbol by
\begin{equation}
\bm{x_\ell} = \bm{F}^H_M \bm{\widetilde{s}_\ell}
\end{equation}
where $\bm{x_\ell}\in\mathcal{C}^{N\times 1}$ is the time-domain samples of the OFDM symbol generated from the $\ell$-th block; $\bm{F}_M\in\mathcal{C}^{M\times M}$ denotes the $M$-dimensional discrete Fourier transform (DFT) matrix, hence $\bm{F}^H_M$ is the inverse DFT (IDFT) matrix; $\bm{\widetilde{s}_\ell}\in\mathcal{C}^{M\times 1}$ is a vector of $M$ complex symbols given by
\begin{equation}
\bm{\widetilde{s}_\ell}[m]=
\begin{cases}
\bm{s_\ell}[n], & \text{if $m=k_n$}, \\
0, & \text{otherwise}.
\end{cases}
\end{equation}
Note that the average power of $\bm{s_{\ell}}$ is $2P$ and the average power of $\bm{\widetilde{s}_{\ell}}$ and $\bm{x_{\ell}}$ is $\frac{2N}{M}P$.

Next, we add cyclic prefix (CP) to the OFDM symbols, yielding $\bm{\widetilde{x}_\ell}\in\mathcal{C}^{(M+L_{\text{cp}})\times 1}$, where $L_{\text{cp}}$ is the CP length. After pulse shaping, the baseband continuous-time signal can be written as
\begin{equation}
    x(t)=\sum_{\ell=0}^{L-1}\sum_{k=0}^{M+L_{\text{cp}}-1}\bm{\widetilde{x}_\ell}[k] p(t-kT-\ell T_{\text{OFDM}}),
\end{equation}
in which $T$ is the baseband sampling period (i.e., the baseband baud rate is $1/T$); $T_{\text{OFDM}}=(M+L_{\text{CP}})T$ is the OFDM symbol duration; and $p(t)$ is the root-raised-cosine (RRC) pulse with a roll-off factor $\beta$.

Finally, the passband signal can be constructed by
\begin{equation}
    x_{\text{RF}}(t)=\text{Re}\left\{x(t)e^{j2\pi f_c t} \right\},
\end{equation}
where $\text{Re}\{*\}$ denotes the real component of a given signal and $f_c$ is the carrier frequency.

After passing through the wireless channel, the received signal is given by
\begin{equation}
r_{\text{RF}}(t)= h(t) \otimes x_{\text{RF}}(t)+w(t),
\end{equation}
where $h(t)$ is the real channel response function; $\otimes$ denotes the linear convolution operation; and $w(t)$ is additive white Gaussian noise (AWGN) with a double-sided power spectral density of $N_0$.
The transmit signal-to-noise ratio (SNR) is defined as $\eta\triangleq P/N_0$.
To study the PAPR of DL-enabled semantic communications, this paper focuses on the AWGN channel case with $h(t)=1$. The extensions to fading channels are straightforward.

As shown in Fig.~\ref{fig:model}, the receiver down converts $r_{\text{RF}}(t)$ to the baseband, matched filters and samples the baseband signal; performs OFDM demodulation, IQ demapping, and DeepJSCC decoding (or deep channel decoding when the source message is a sequence of i.i.d. bits) to recover the transmitted message.
Denote by $\widehat{\bm{s}}_{\text{msg}}$ the reconstructed source message. The communication system is trained end-to-end to minimize a loss function $\mathcal{L}$ between ${\bm{s}}_{\text{msg}}$ and $\widehat{\bm{s}}_{\text{msg}}$. Depending on whether ${\bm{s}}_{\text{msg}}$ is continuous or discrete, typical loss functions are the mean-squared error (MSE) and the cross-entropy loss.

\section{PAPR of Discrete-time Analog Transmission}\label{sec:III}
Unlike prior works that emphasize message reconstruction performance, this paper focuses on the PAPR perspective of DL-enabled semantic communications with DTAT.
In actuality, achieving high reconstruction performance and low PAPR is a trade-off, because a low-PAPR transmitted signal offers a limited degree of freedom for message reconstruction.

DeepJSCC-enabled semantic communications often exhibit much better message reconstruction performance than digital communications \cite{Deniz2022,JSCC2019}, but the PAPR of the transmitted signal is also much larger.
A natural question that arises is whether the reconstruction gains of DeepJSCC come from the higher-PAPR signal. Said in another way, can the excellent reconstruction performance be retained if the PAPR of the transmitted signal is bounded to levels permissible for practical implementations? 
Our main objective in this paper is to answer the above questions and characterize the trade-off between the reconstruction performance and the PAPR performance of DeepJSCC with DTAT.

We measure the PAPR performance of a communication system by the PAPR of the passband signal $x_{\text{RF}}(t)$:
\begin{equation}\label{eq:rho}
    \rho = \frac{\max |x_{\text{RF}}(t)|^2}{\mathbb{E}\left[|x_{\text{RF}}(t)|^2\right]}.
\end{equation}

The complementary cumulative distribution function (CC\-DF) of $\rho$ is defined as $\Pr(\rho>\Gamma)$ for a threshold $\Gamma$ (i.e., the tail distribution).
As a PAPR performance indicator, we shall use the $99.9$-percentile PAPR, denoted by $\Gamma_{-3}$, where $\Pr(\rho>\Gamma_{-3})=10^{-3}$.

\subsection{PAPR Analysis}\label{sec:IIIA}
For a well-trained DeepJSCC-based communication system, the PAPR of the passband signal is non-trivial to derive in general, because it depends on the source message, the channel distribution, and the learning process, which are elusive to characterize analytically.
A simple analysis is given below by assuming that the DeepJSCC coded symbols $\bm{s}_{\text{enc}}$ follow i.i.d. Gaussian distributions.

Under this assumption, $\bm{s}_{\text{enc}}$ is a multivariate Gaussian random vector.
After power normalization, the vector $\bm{s}_{\text{norm}}$ follows $\bm{s}_{\text{norm}}\sim\mathcal{N}(\bm{0},P\bm{I})$, where $\bm{0}$ is an all-zero vector and $\bm{I}$ is the identity matrix.
After IQ mapping, partition, and subcarrier mapping, each block of complex symbols $\bm{s_\ell}\sim\mathcal{CN}(\bm{0},2P\bm{I})$ and $\bm{\widetilde{s}_\ell}\sim\mathcal{CN}(\bm{0},2P\bm{\Lambda})$, where $\bm{\Lambda}$ is an $M$-dimensional diagonal matrix. In particular, the $m$-th diagonal element of $\bm{\Lambda}$ is $1$ if $m\in\{k_1,k_2,k_3,...,k_N\}$ and $0$ otherwise.
After IDFT, the time-domain OFDM samples follow $\bm{x_\ell}\sim\mathcal{CN}(\bm{0},2P\bm{F}^H\bm{\Lambda}\bm{F})$. In particular, each element of $\bm{x_\ell}$ follows $\bm{x_\ell}[m]\sim\mathcal{CN}(\bm{0},\frac{2N}{M}P)$ and $\|\bm{x_\ell}[m]\|^2_2\sim \exp(\frac{M}{2NP})$, $\forall m$. The average power of $\bm{x_\ell}$ is $\frac{2N}{M}P$, hence, the PAPR of $\bm{x_\ell}$ is given by
\begin{equation}
 \rho(\bm{x_\ell}) = \frac{M}{2NP}\max_m \|\bm{x_\ell}[m]\|^2_2.
\end{equation}

The CCDF of $\rho(\bm{x_\ell})$ can be approximated by
\begin{eqnarray}
\Pr\left(\rho(\bm{x_\ell})>\Gamma\right) \hspace{-0.2cm}&=&\hspace{-0.2cm}
\Pr\left(\max_m \|\bm{x_\ell}[m]\|^2_2>\frac{2NP}{M}\Gamma\right) \nonumber\\
\hspace{-0.2cm}&\approx&\hspace{-0.2cm}
1-\prod_{m=0}^{M-1} \Pr\left( \|\bm{x_\ell}[m]\|^2_2 < \frac{2NP}{\kappa_1 M}\Gamma \right) \nonumber\\
\hspace{-0.2cm}&=&\hspace{-0.2cm}
1- \left(1-e^{-\frac{\Gamma}{\kappa_1}} \right)^M.
\end{eqnarray}
where the approximation comes from the assumption that the elements of $\bm{x_\ell}$ are independent, while they are in fact correlated since the covariance of $\bm{x_\ell}$ is $2P\bm{F}^H\bm{\Lambda}\bm{F}$.
To compensate for the loss of this independent assumption, a hyperparameter $\kappa_1$ is introduced. It is worth noting that the independent assumption is valid when $N=M$, since $\bm{x_\ell}\sim\mathcal{CN}(0,2P\bm{I})$, in which case we set $\kappa_1=1$.

For the passband signal $x_{\text{RF}}(t)$, the CCDF of $\rho$ in \eqref{eq:rho} can be approximated by that of oversampled baseband signal $\bm{x_\ell}$ \cite{PAPRapprox}, yielding,
\begin{equation}\label{eq:rho_approx}
\rho \approx 1- \left(1-e^{-\frac{\Gamma}{\kappa_1}} \right)^{\kappa_2 M},
\end{equation}
where  $\kappa_2$ is another hyperparameter. Both $\kappa_1$ and $\kappa_2$ can be found by parametric fitting.

% For IFDMA, the PAPR of the time-domain OFDMA samples $\bm{x_\ell}$ is the same as that of $\bm{s_\ell}$, according to \eqref{eq:IFDMAconstruction}. Since $\bm{s_\ell}\sim\mathcal{CN}(\bm{0},2P\bm{I})$, we have $\|\bm{s_\ell}[n]\|^2_2\sim \exp(\frac{1}{2P})$, $\forall n$, and
% \begin{equation}
%  \rho(\bm{x_\ell})  = \rho(\bm{s_\ell}) = \frac{1}{2P}\max_n \|\bm{s_\ell}[n]\|^2_2.
% \end{equation}

% The CCDF of $\rho(\bm{x_\ell})$ is given by
% \begin{eqnarray}
% \Pr\left(\rho(\bm{x_\ell})>\Gamma\right) \hspace{-0.2cm}&=&\hspace{-0.2cm}
% \Pr\left(\max_n \|\bm{s_\ell}[n]\|^2_2>2P\Gamma\right) \nonumber\\
% \hspace{-0.2cm}&=&\hspace{-0.2cm}
% 1- \left(1-e^{-\Gamma} \right)^N.
% \end{eqnarray}
% The CCDF of $\rho$ in \eqref{eq:IFDMAconstruction} for the passband signal $x_{\text{RF}}(t)$ can be approximated by that of the baseband signal $\bm{x_\ell}$ \cite{?}, yielding,
% \begin{equation}\label{eq:rhoIFDMA}
% \rho \approx 1 - \left(1-e^{-\Gamma} \right)^{\kappa N}.
% \end{equation}
% where $\kappa$ is a constant that can be found by parametric fitting.

\subsection{PAPR reduction}\label{sec:IIIB}
A low PAPR is essential for achieving high power efficiency at the transmitter. This paper explores three PAPR reduction techniques and characterizes the trade-off between message reconstruction and PAPR with these techniques.

\begin{figure*}[t]
  \centering
  \includegraphics[width=0.85\linewidth]{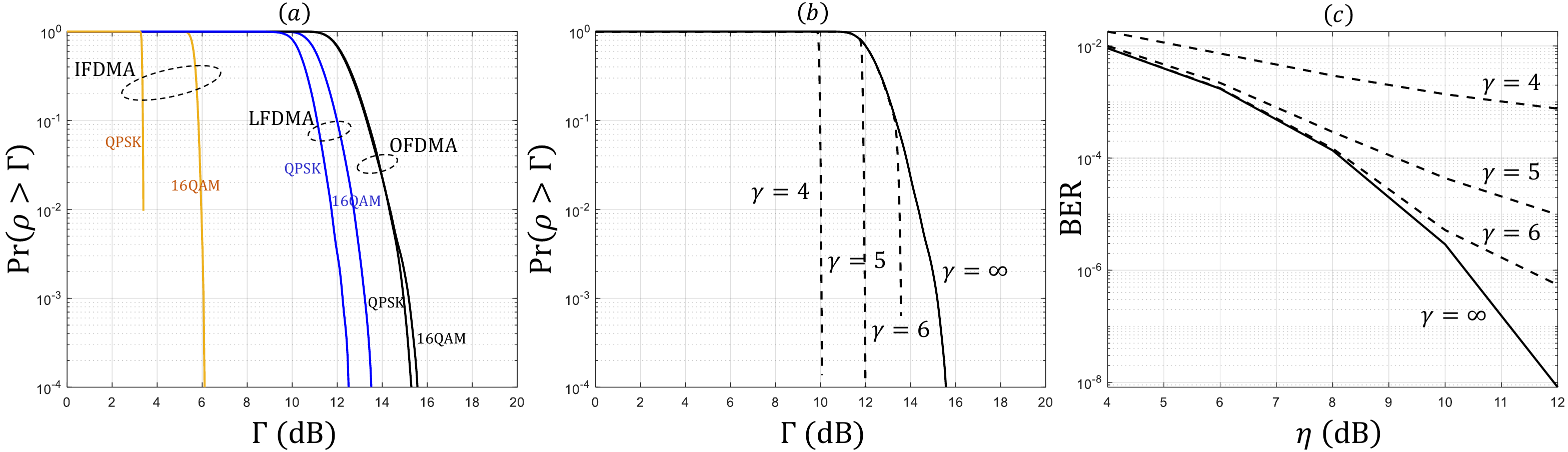}\\
  \caption{PAPR reduction in digital communications. The system configurations are given in Table~\ref{tab:1}. (a) PAPR performance of linearly precoded OFDMA; (b) PAPR performance with clipping, where the modulation is 16QAM, no channel code is considered, and $\gamma=\infty$ means no clipping. (c) Impact of clipping on the BER performance.}
\label{fig:sim_digital}
\end{figure*}

\subsubsection{Linearly precoded OFDMA}
In digital communications, an efficient scheme to reduce the PAPR of the transmitted signal without compromising the BER performance is linearly precoded OFDMA (a.k.a., single carrier FDMA)\cite{SCFDMA,shaoIFDMA}, which is standardized in both 4G LTE and 5G NR for uplink low-PAPR transmission.

Linearly-precoded OFDMA generates low-PAPR waveforms by changing the signal modulation and subcarrier allocation schemes. Specifically, with OFDMA, the $N$ allocated subcarriers are randomly distributed over $\mathcal{M}$ and a block of complex symbols $\bm{s_\ell}\in\mathcal{C}^{N\times 1}$ is directly mapped onto the $N$ subcarriers, as explained in Section \ref{sec:II}.
With linearly-precoded OFDMA, on the other hand, we first DFT-precode $\bm{s_\ell}$, yielding
\begin{equation}
\bm{s^\prime_\ell}= \bm{F}_N \bm{s_\ell},
\end{equation}
and then map the precoded symbols $\bm{s^\prime_\ell}$ onto the subcarriers. In particular, the $N$ allocated subcarriers are localized (contiguous) or interleaved (evenly-spaced) over the spectrum. With localized and interleaved subcarrier allocations, the corresponding linear precoded OFDMA systems are called LFDMA and IFDMA, respectively.

In digital communications, the BER performances of OFDMA, LFDMA, and IFDMA are exactly the same, but the IFDMA waveform exhibits the lowest PAPR. The reason is as follows. Consider a block of discrete constellations $\bm{s_\ell}$. After $N$-DFT precoding, interleaved subcarrier mapping, and $M$-IDFT modulation, it can be shown that the resulting signal is given by
\begin{equation}\label{eq:IFDMAconstruction}
\bm{x_\ell}[m]=\frac{N}{M} e^{j \frac{2\pi m}{M}} \bm{s_\ell}[m~\text{mod}~N], 
\end{equation}
for $m=0,1,2,..., M-1$. That is, $\bm{x_\ell}$ can be constructed directly from $\bm{s_\ell}$ by simple repetition and frequency upshifting. This implies that the IFDMA waveform exhibits nearly the same PAPR properties as the discrete constellations $\bm{s_\ell}$, and hence, has much lower PAPR than that of LFDMA and OFDMA.

A visual illustration is given in Fig.~\ref{fig:sim_digital}(a), where we simulate a digital communication system with QPSK and 16QAM modulations. As can be seen, with 16QAM, the 99.9-percentile PAPR $\Gamma_{-3}$ of OFDMA is improved by 2dB and 9 dB with LFDMA and IFDMA, respectively. The performance gains are even larger when lower-order modulations are used. For QPSK, the $\Gamma_{-3}$ gains of LFDMA and IFDMA over OFDMA are up to 3dB and 11.5dB, respectively.

% With linearly precoded OFDMA, the receiver will perform an additional $N$-IDFT after $M$-DFT in Fig.~\ref{fig:model}. 

\subsubsection{Clipping}
Clipping the large amplitude of $x_{\text{RF}}(t)$ is a straightforward PAPR reduction scheme \cite{clip}. Specifically, let $\bar{x}_{\text{RF}}=\mathbb{E}\left(\left|x_{\text{RF}}(t)\right|\right)$ be the average amplitude of $x_{\text{RF}}(t)$. We clip $x_{\text{RF}}(t)$ such that the magnitude of the clipped signal does not exceed a threshold $\gamma \bar{x}_{\text{RF}}$, where $\gamma$ is a clipping ratio measuring the severity of clipping. The clipped signal can be written as
\begin{equation}\label{eq:clip1}
x_{\text{clip}}(t)=\begin{cases}
x_{\text{RF}}(t), & \text{if}~\left| x_{\text{RF}}(t) \right|\leq \gamma \bar{x}_{\text{RF}}; \\
\gamma \bar{x}_{\text{RF}}, & \text{if}~\left| x_{\text{RF}}(t) \right|> \gamma \bar{x}_{\text{RF}}.
\end{cases}
\end{equation}

Fig.~\ref{fig:sim_digital}(b) presents the PAPR reduction performance of clipping in digital OFDMA communication systems, where the modulation is 16QAM and various clipping ratios are considered. As shown, increasingly lower PAPR can be obtained as we decrease $\gamma$. When $\gamma=5$, for example, the 99.9-percentile PAPR is improved by $3.1$ dB.
Clipping, however, causes both out-of-band radiation and in-band inter-carrier interference (ICI). The out-of-band radiation can be addressed by filtering according to the spectrum mask. The in-band ICI, on the other hand, leads to an inevitable BER loss since the orthogonality among subcarriers is destroyed. Fig.~\ref{fig:sim_digital}(c) shows the impact of clipping on the BER. When $\gamma=5$, the BER performance deteriorates for $2.7$ dB to achieve a BER of $10^{-5}$.

\subsubsection{PAPR loss}
Since our goal is to minimize both the reconstruction error and the PAPR of $x_{\text{RF}}(t)$, a natural idea is to add a PAPR loss to the original reconstruction loss \cite{PAPRloss1}, and minimizing 
\begin{equation}
\mathcal{L}'= \mathcal{L}+\lambda\mathbb{E}[\rho],
\end{equation}
where $\lambda$ is a hyperparameter.
With this new loss function, the communication system learns to minimize both metrics simultaneously, where $\lambda$ determines the operating points on the trade-off curve between MSE and PAPR.

\begin{figure*}[t]
  \centering
  \includegraphics[width=0.7\linewidth]{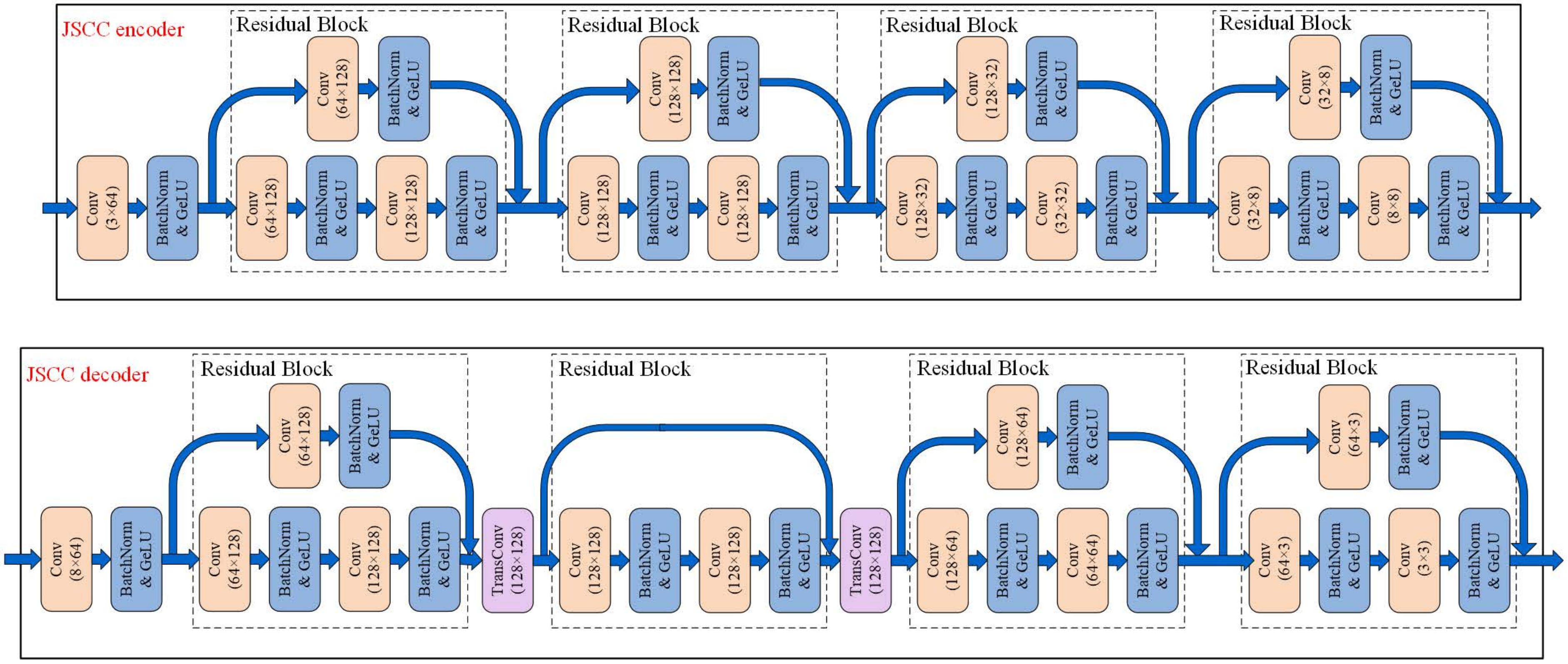}\\
  \caption{The DeepJSCC encoder and decoder architectures  for semantic image transmission.}
\label{fig:DNN}
\end{figure*}

\section{Simulation Results}\label{sec:IV}

\begin{table}[t]
\renewcommand*{\arraystretch}{0.9}
    \caption{Hyperparameter settings.}
    \label{tab:1}
    \centering
\begin{tabular}{llll}
\toprule
                    & Hyper parameters & Symbols          & Values \\ \midrule
\multirow{8}{*}{System}   & Bandwidth ratio         & $R$                 & $1/12$    \\
                    & Total number of subcarriers         & $M$                 & $128$    \\
                    & Number of allocated subcarriers     & $N$                 & $64$     \\
                    & Length of CP                        & $L_{\text{cp}}$     & $16$     \\
                    & Roll-off factor of RRC                        & $\beta$     & $0.5$     \\
                    & Carrier frequency                   & $f_c$               & $25$MHz  \\
                    & Baseband baud rate                  & $1/T$               & $1$MHz   \\
                    & Baseband oversampling rate          &                     & $10$MHz  \\\midrule
\multirow{5}{*}{Learning}   & Number of training epochs           &                     & $100$    \\
                    & Batch size                          &                     & $256$    \\
                    & Learning rate                       &                     & $10^{-3}$   \\
                    & Weight decay                        &                     & $5\times 10^{-3}$   \\
                    & Optimizer                           &                     & adamW  \\
\bottomrule
\end{tabular}
\end{table}

\begin{figure}
     \centering
     \begin{subfigure}
         \centering
         \includegraphics[width=0.7\columnwidth]{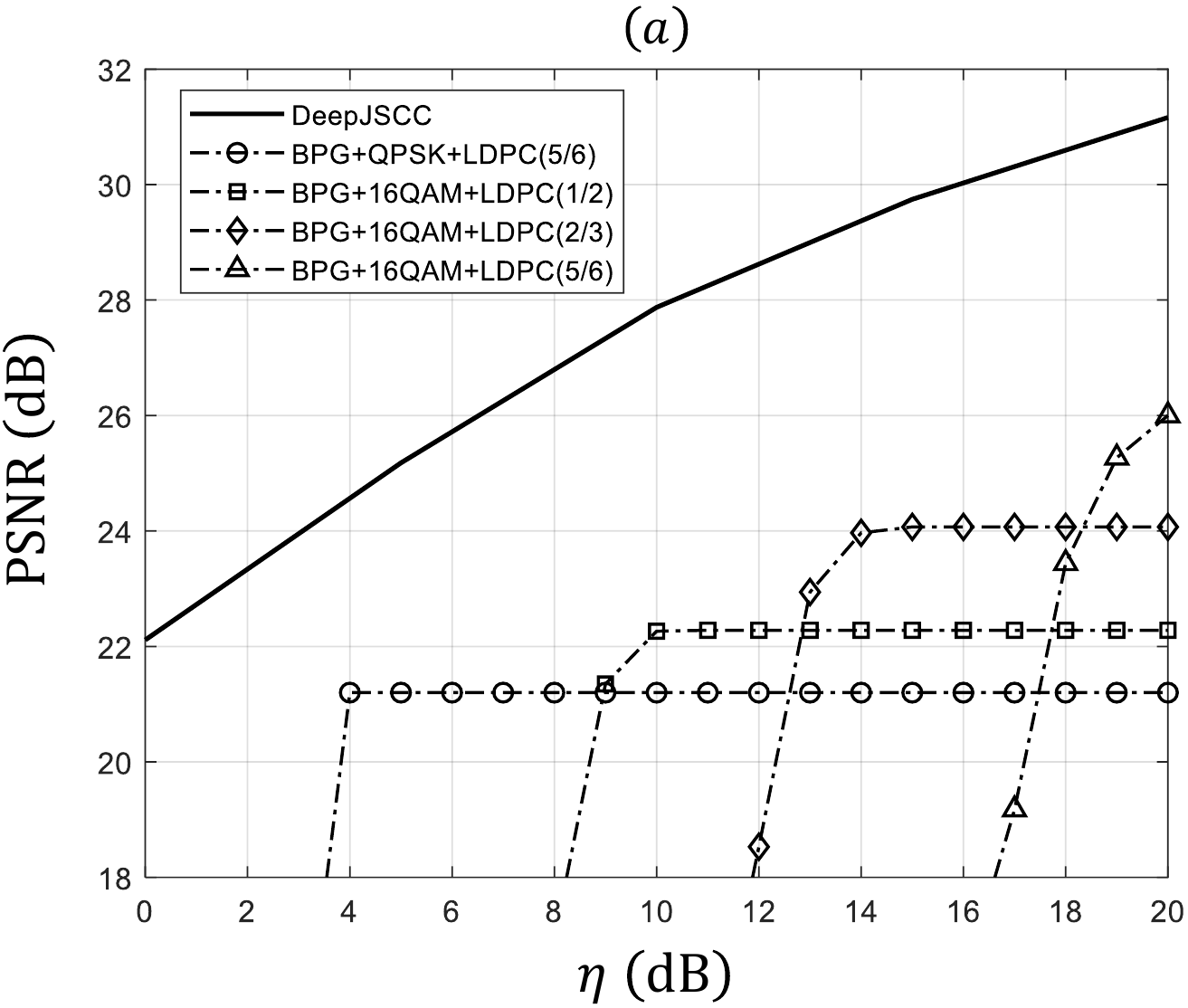}
     \end{subfigure}
    %  \vspace{0.1cm}
     \begin{subfigure}
         \centering
         \includegraphics[width=0.7\columnwidth]{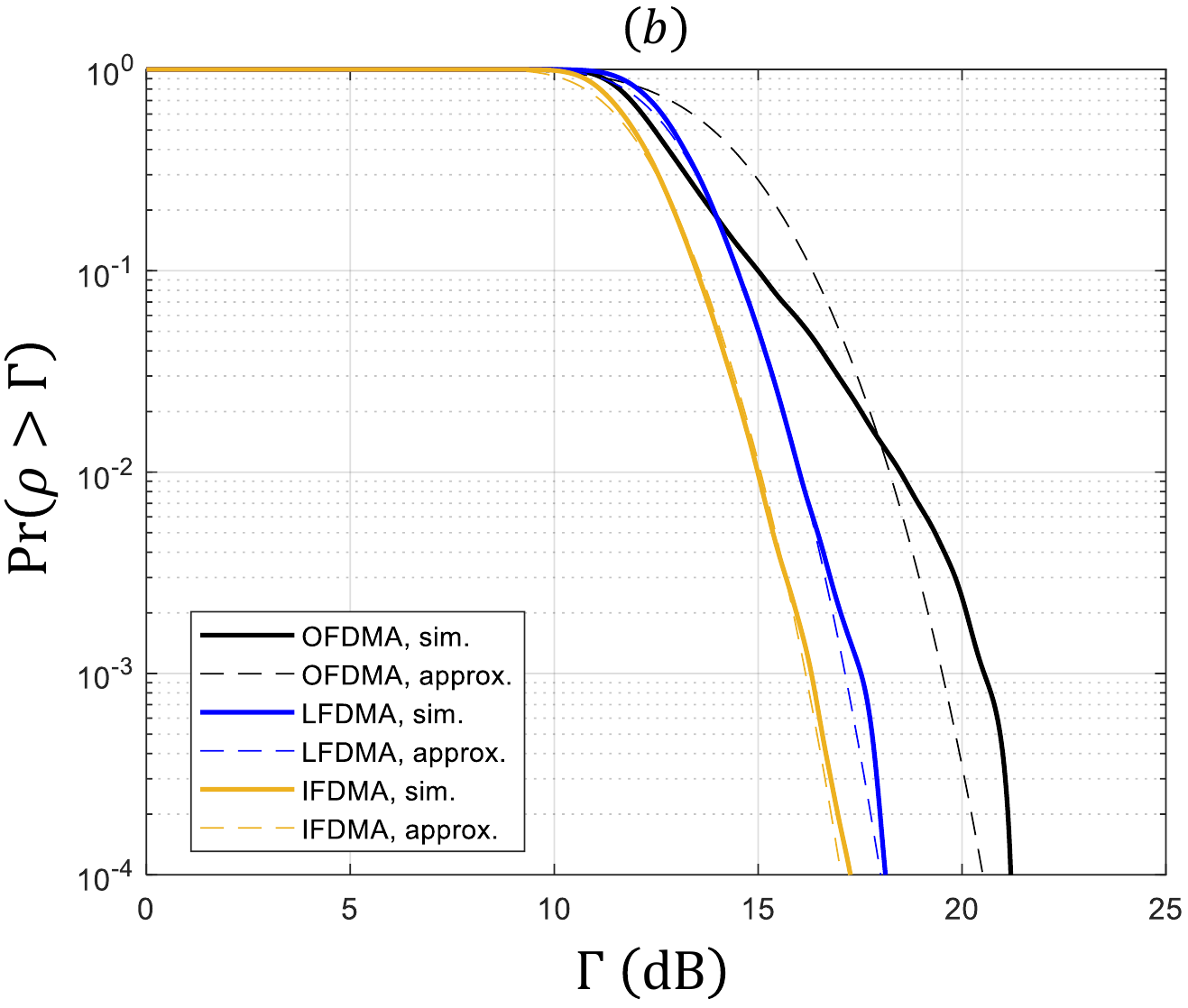}
     \end{subfigure}
  \caption{PSNR and PAPR performances of DeepJSCC. (a) PSNR under various SNR $\eta$; (b) CCDF of PAPR for the model trained at $\eta=10$ dB. The dashed curves are plotted using \eqref{eq:rho_approx}.}
\label{fig:sim1}
\end{figure}

\subsection{Transmission of CIFAR10 images}
This section analyzes the PAPR performance of DL-enable semantic communications considering the wireless transmission of CIFAR10 images. In this application, the source messages are assumed to be colored images from the CIFAR10 dataset. To retain the spatial feature of images, we denote a source message by a three-dimensional matrix $\bm{S}_{\text{msg}}\in\mathcal{R}^{32\times 32\times 3}$. If reshaped to a vector $\bm{s}_{\text{msg}}\in\mathcal{R}^{L_s\times 1}$ as defined in Section \ref{sec:II}, we have $L_s=32\times 32\times 3$. The DeepJSCC encoder is designed to be a residual convolutional neural network (CNN) with convolutional layers, batch normalization (BN) layers, and Gaussian error linear unit (GeLU) activation functions. The detailed network architecture is depicted in Fig. \ref{fig:DNN}. For a given compression ratio $R$, the source image is transformed to a feature matrix $\bm{S}_{\text{enc}}\in\mathcal{R}^{8\times 8\times 96R}$ and then reshaped to a vector $\bm{s}_{\text{enc}}\in\mathcal{R}^{L_e\times 1}$ with $L_e=8\times 8\times 96R$.

Following the signal flow defined in Section \ref{sec:II}, we process $\bm{s}_{\text{enc}}$ by power normalization, IQ mapping, and Tx modulation at the transmitter; Rx demodulation and IQ demapping at the receiver, after which we obtain the noisy feature matrix $\widehat{\bm{S}}_{\text{enc}}\in\mathcal{R}^{8\times 8\times 96R}$. Then, we feed $\widehat{\bm{S}}_{\text{enc}}$ into the DeepJSCC decoder to reconstruct the source image $\widehat{\bm{S}}_{\text{msg}}\in\mathcal{R}^{32\times 32\times 3}$. The DeepJSCC decoder is designed to be a residual network with transposed convolutional layers, BN layers, and GeLU activation functions, as depicted in Fig. \ref{fig:DNN}.

The communication goal is image reconstruction and we aim to maximize the PSNR of the reconstructed image:
\begin{equation}
\text{PSNR}= 10\log \frac{L_s}{\mathbb{E}\|\widehat{\bm{S}}_{\text{msg}}-{\bm{S}}_{\text{msg}}\|^2_F},
\end{equation}
where $\|*\|_F$ denotes the Frobenius norm of a matrix. To this end, the loss function is chosen to be the MSE loss
\begin{equation}\label{eq:PSNRloss}
\mathcal{L}= \mathbb{E}\left[\frac{1}{L_s}\|\widehat{\bm{S}}_{\text{msg}}-{\bm{S}}_{\text{msg}}\|^2_F\right].
\end{equation}

As a baseline, we train the semantic image transmission system with OFDMA in an end-to-end fashion to minimize \eqref{eq:PSNRloss} under various transmit SNRs $\eta$ from $0$ dB to $20$ dB. 
The system and learning configurations are summarized in Table~\ref{tab:1} unless otherwise specified.
The PSNR versus $\eta$ performance is presented in Fig.~\ref{fig:sim1}(a), where the benchmarks are the traditional digital communication systems with better portable graphics (BPG) image compression, low-density parity-check code (LDPC) with different rates, and various modulation schemes.
As shown, the PSNR of DeepJSCC outperforms that of digital communication.
At an SNR of $10$ dB, DeepJSCC achieves a PSNR of $27.83$, while the best PSNR achieved by digital communication is $22.26$.

Let us further evaluate the PAPR performance of both systems.
For digital communication, the best PSNR at $\eta=10$ dB is achieved by BPG, 16QAM, and LDPC with a $1/2$ rate, the 99.9-percentile PAPR of which is about $15$ dB with OFDMA.
On the other hand, the 99.9-percentile PAPR of DeepJSCC is up to $21$ dB with OFDMA.
This indicates that, despite the excellent PSNR performance, PAPR is a limitation for the practical implementation of DeepJSCC-enabled semantic communication. PAPR redution techniques must be introduced.

% In actuality, achieving high PSNR and low PAPR is a trade-off since a low PAPR leads to less freedom to attain a large PSNR.
% A natural question that arises is whether the PSNR gains of semantic communication come from the higher PAPR. Said in another way, can the excellent PSNR performance of semantic communication retain if the PAPR of a transmitted signal is bounded? In the following, we answer this question and characterize the trade-off between PAPR and PSNR with three PAPR reduction techniques, i.e., linearly precoded OFDMA, clipping, and PAPR loss. 

% digital + 16QAM
% OFDMA: 15 dB
% LFDMA: 13 dB
% IFDMA: 6 dB

% analog + 16QAM
% OFDMA: 21 dB
% LFDMA: 17.5 dB
% IFDMA: 16.5 dB

% Sim improvements:
% 1. increase epochs such that we outperform benchmark 28.5?
% 2. PAPR loss, instead of db, use times
% 3. transmit 5 times for each img, (5 can be larger depending on the GPU size).
% 4. geLu - ReLu - Leaky Relu

\subsection{DeepJSCC with PAPR reduction}
To reduce the PAPR of the continuous-amplitude signal, this section applies the three PAPR reduction techniques discussed in Section \ref{sec:IIIB} and studies the PSNR versus PAPR trade-off for DeepJSCC-enabled semantic communications.

\subsubsection{Linearly-precoded OFDMA}
% Linearly precoded OFDMA can alleviate the large PAPR to some extent, but the gain is not as large as digital comm. Because of the constellation-free nature of semantic communications.
We first study the PAPR reduction capability of linearly precoded OFDMA.
At an SNR of $10$ dB, the CCDF of PAPR is presented in Fig.~\ref{fig:sim1}(b). We have two main observations,
\begin{itemize}
\item With linearly-precoded OFDMA, the PAPR reduction in DeepJSCC is not as significant as that in digital communication -- the 99.9-percentile PAPR of OFDMA is only reduced by $3.5$ dB and $4.5$ dB with LFDMA and IFDMA, respectively. This observation is not surprising, because DeepJSCC outputs continuous amplitude signals as opposed to discrete constellations.
\item With IFDMA, both DeepJSCC and digital communications exhibit the lowest PAPR. The 99.9-percentile PAPR of DeepJSCC is $10.5$ dB larger than that of digital communication with 16QAM modulation.
\end{itemize}

\begin{figure}[t]
  \centering
  \includegraphics[width=0.7\linewidth]{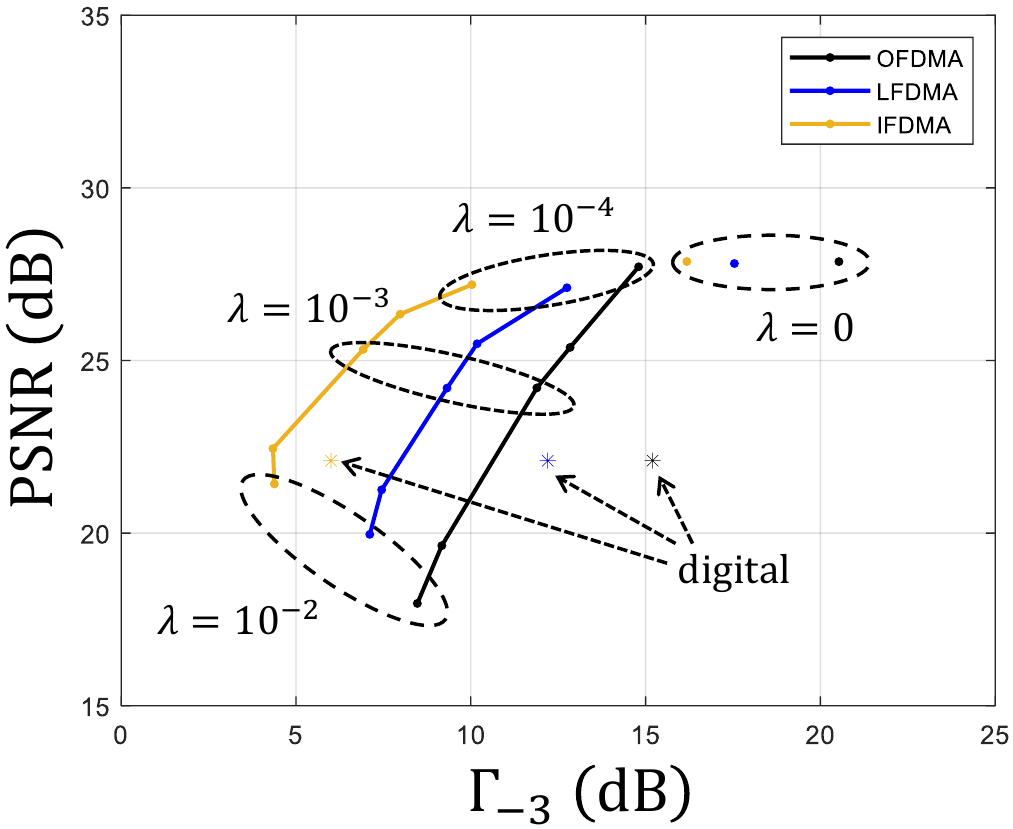}\\
  \caption{PSNR versus PAPR ($\Gamma_{-3}$) of DeepJSCC with PAPR loss.}
\label{fig:PAPRloss}
\end{figure}

\subsubsection{PAPR loss}
To further reduce the PAPR, we introduce an additional PAPR loss to the PSNR loss. The new loss function is given by
\begin{equation}\label{eq:PSNRPAPRloss}
\mathcal{L}'= \mathbb{E}\left[\frac{1}{L_s}\|\widehat{\bm{S}}_{\text{msg}}-{\bm{S}}_{\text{msg}}\|^2_F\right]+\lambda\mathbb{E}[\rho].
\end{equation}

With different $\lambda$, DeepJSCC can balance PSNR and PAPR.
To illustrate the trade-off, we plot the ROC curve in Fig.~\ref{fig:PAPRloss}.
The ROC space is defined by the 99.9-percentile PAPR $\Gamma_{-3}$ (in dB) and PSNR (in dB) as the $x$ and $y$ axis.
The performance of a DeepJSCC-based semantic communication system can be represented by one point in the ROC space and the best system yields a point in the upper left corner of the space, in which case the PSNR is maximized and the PAPR is minimized.

We have two main observations from Fig.~\ref{fig:PAPRloss}.
\begin{itemize}
\item With the increase in $\lambda$, PAPR weighs more than MSE in \eqref{eq:PSNRPAPRloss}. As a result, the well-trained DeepJSCC model exhibits lower PAPR but also lower PSNR.
\item IFDMA exhibits a better trade-off than LFDMA and OFDMA, thanks to its simpler signal structure at the transmitter. That is, with IFDMA, it is easier to learn a communication system with low PAPR and high PSNR by minimizing \eqref{eq:PSNRPAPRloss}. When $\lambda=10^{-4}$, for example, IFDMA, LFDMA, and OFDMA yield almost the same PSNR performance, but the 99.9-percentile PAPR of IFDMA is about $2.5$ dB and $4.9$ dB less than that of LFDMA and OFDMA, respectively.
\end{itemize}

\subsubsection{Clipping}
\begin{figure}
     \centering
     \begin{subfigure}
         \centering
         \includegraphics[width=0.72\columnwidth]{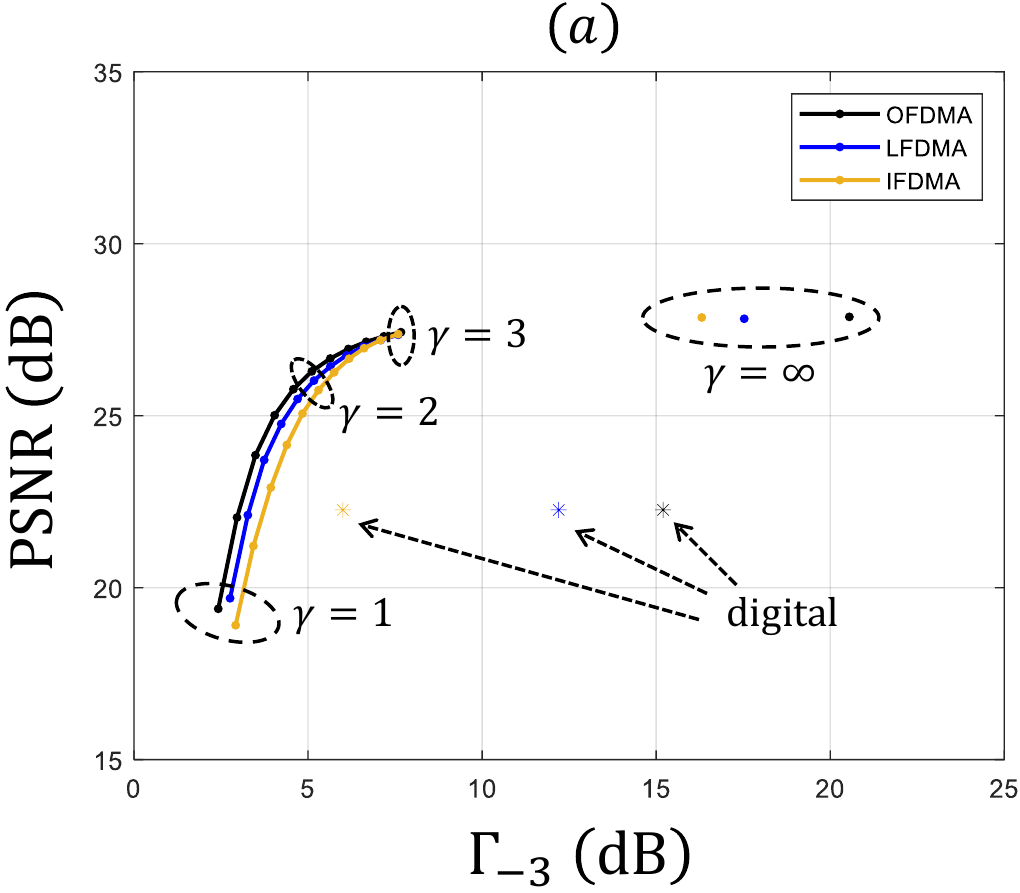}
     \end{subfigure}
    %  \vspace{0.1cm}
     \begin{subfigure}
         \centering
         \includegraphics[width=0.72\columnwidth]{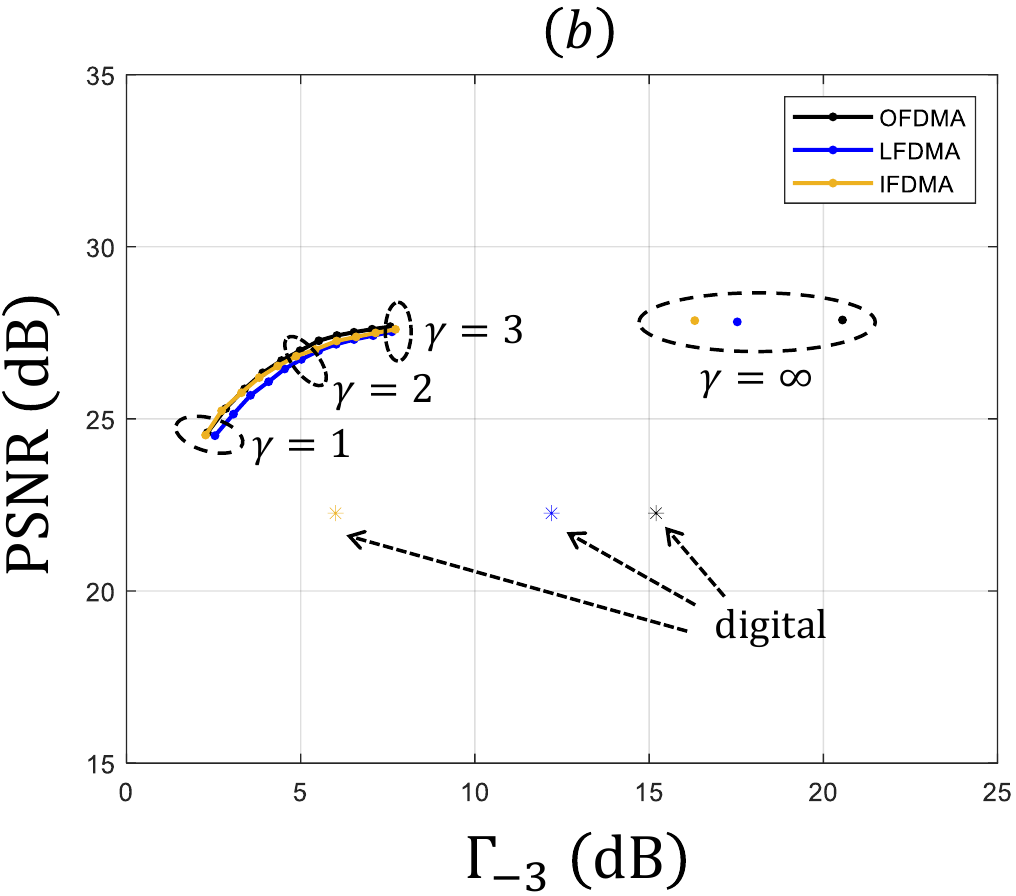}
     \end{subfigure}
  \caption{PSNR versus PAPR ($\Gamma_{-3}$) of DeepJSCC with clipping: (a) clipping without retraining; (b) clipping with retraining.}
\label{fig:clip}
\end{figure}

Finally, we evaluate the performance of clipping when applied to DeepJSCC-based image transmission.
Fig. \ref{fig:clip}(a) characterizes the PSNR and PAPR trade-off when we clip $x_{\text{RF}}(t)$ of a well-trained DeepJSCC encoder according to \eqref{eq:clip1}. Note that the signal after clipping has to be re-normalized such that the power of $\widetilde{x}_{\text{clip}}(t)$ is the same as that of $x_{\text{RF}}(t)$, that is,
\begin{equation}
\widetilde{x}_{\text{clip}}(t)=\sqrt{\frac{x^2_{\text{RF}}(t)}{x^2_{\text{clip}}(t)}} x_{\text{clip}}(t).
\end{equation}

As shown in Fig. \ref{fig:clip}(a), clipping can reduce the PAPR by a large margin without compromising too much PSNR performance. When $\gamma=3$, the 99.9-percentile PAPR of OFDMA is reduced by $13.5$ dB, while the PSNR only deteriorates by $0.45$ dB.  When $\gamma=2$, the 99.9-percentile PAPR of OFDMA is reduced by $16$ dB, while the PSNR only deteriorates by $1.58$ dB. In other words, the DeepJSCC decoder is robust to the severe ICI caused by clipping, and a good trade-off between PSNR and PAPR can be achieved. 

Moreover, we can incorporate the clipping operation into the training of the communication system and further enhance the robustness of the trained system to clipping. To this end, we implement a trainable clipping operation by
\begin{equation}
\widetilde{x}_{\text{clip}}(t)=x_{\text{RF}}(t)\left(1-\frac{\text{ReLU}\left(|x_{\text{RF}}(t)|- \gamma \bar{x}_{\text{RF}}\right)}{|x_{\text{RF}}(t)|+\varepsilon}   \right),
\end{equation}
where $\varepsilon=10^{-8}$ is a small constant for numerical stability.
Fig. \ref{fig:clip}(b) characterizes the PSNR and PAPR trade-off when clipping is incorporated into the training phase. As shown,
\begin{itemize}
\item Compared with Fig. \ref{fig:clip}(a), DeepJSCC is even more robust to small clipping ratios. When $\gamma=1$, the 99.9-percentile PAPR of OFDMA is reduced by $18.5$ dB, while the PSNR loss is only $3.27$ dB.
\item There is no remarkable difference between OFDMA and linearly-precoded OFDMA after clipping. Thus, the additional DFT (IDFT) operation at the transmitter (receiver) is unnecessary.
\item Suppose the 99.9-percentile PAPR of 16QAM and IFDMA in digital communications is an acceptable PAPR performance, i.e., $\Gamma_{-3}=6$ dB. With clipping, the PSNR is up to $27.42$ when the target 99.9-percentile PAPR is $6$ dB.
\end{itemize}

\begin{figure}[t]
  \centering
  \includegraphics[width=0.75\linewidth]{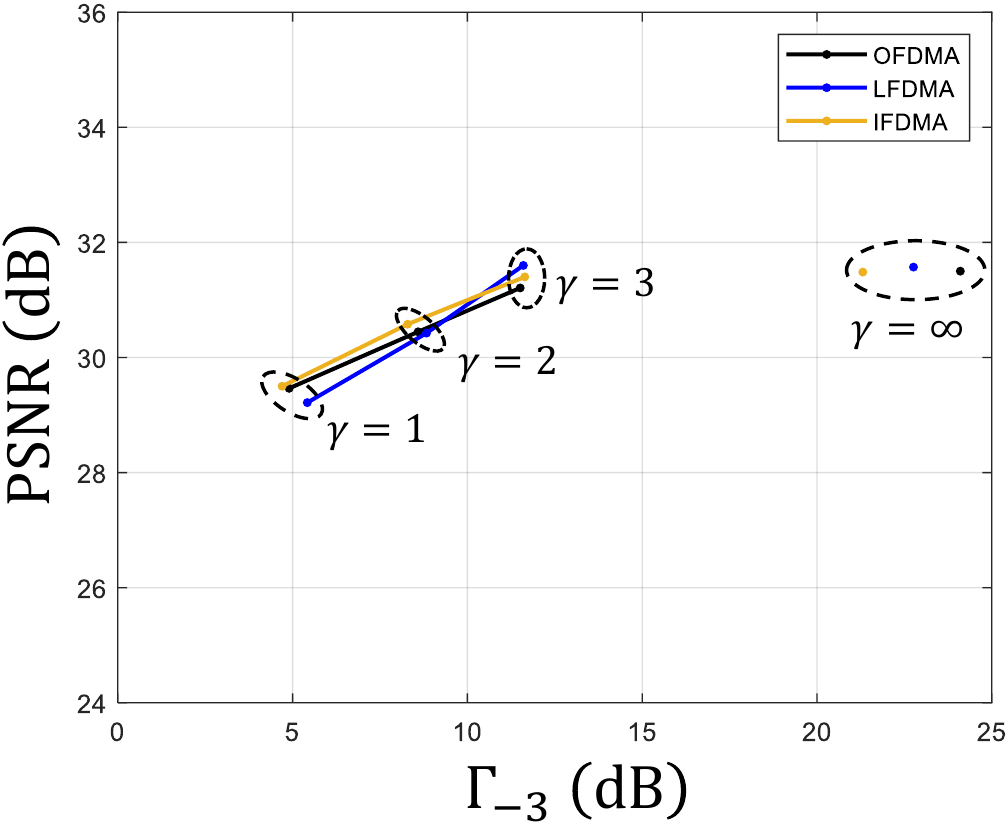}\\
  \caption{PSNR versus PAPR ($\Gamma_{-3}$) of DeepJSCC with clipping (with retraining) on ImageNet.}
\label{fig:ImageNet}
\end{figure}

To confirm the above results, we further extend our experiments to ImageNet and evaluate the impact of clipping with retraining on the PAPR and PSNR performances.
ImageNet consists of 1.2 million high-resolution images.
In the training phase, we randomly sample mini-batches of size $32$ from ImageNet and crop each image to size $128 \times 128$. As the experiments on CIFAR10, the bandwidth ratio is fixed to $1/12$ and each transmission packet consists of $256$ complex symbols ($4$ OFDM symbols). The number of training epochs is fixed to $5$.
In the test phase, we evaluate the trained DeepJSCC model on the Kodak dataset, which consists of $24$ images of size $768 \times 512$. The experimental results are shown in Fig.~\ref{fig:ImageNet}. As can be seen, setting $\gamma=1$ satisfies the target 99.9-percentile PAPR of $6$ dB, while the PSNR degradation is only $2$ dB compared with the no clipping case.

\section{Conclusion}\label{sec:6}
DL-based data-driven approaches are playing an increasingly important role in the physical layer of wireless communications, particularly for semantic communications that have received significant research attention.
% Promising directions include DL-aided modulation, channel coding, JSCC, etc.
An indispensable part of the DL-aided physical-layer communication system design is DTAT, where DL techniques allow the optimization of encoded signals without being limited to fixed finite signal constellations.
DTAT, however, often leads to a high PAPR of the transmitted signal.
The significance of this paper is twofold.
\begin{itemize}
\item We point out the PAPR problem of DTAT in DL-aided wireless communications and provide a passband transceiver that can be used in other applications to evaluate the PAPR performance.
\item We show that the high PAPR of the continuous-amplitude signal obtained by the DeepJSCC encoder in semantic image transmission can be tackled by incorporating clipping into the training process. Our numerical experiments reveal that high-quality image reconstruction and low-PAPR transmission can be achieved simultaneously.
% \item Although DTAT is necessary, we confirm that the high PAPR of the continuous-amplitude signal is noncritical in semantic image transmission. High-quality image reconstruction and low-PAPR transmission can be achieved simultaneously.
\end{itemize}

\appendices

% \section{Experimental setup}\label{sec:AppA}
% \input{AppendixA.tex}

\bibliographystyle{IEEEtran}
\bibliography{References}

\begin{thebibliography}{10}
\providecommand{\url}[1]{#1}
\csname url@samestyle\endcsname
\providecommand{\newblock}{\relax}
\providecommand{\bibinfo}[2]{#2}
\providecommand{\BIBentrySTDinterwordspacing}{\spaceskip=0pt\relax}
\providecommand{\BIBentryALTinterwordstretchfactor}{4}
\providecommand{\BIBentryALTinterwordspacing}{\spaceskip=\fontdimen2\font plus
\BIBentryALTinterwordstretchfactor\fontdimen3\font minus
  \fontdimen4\font\relax}
\providecommand{\BIBforeignlanguage}[2]{{%
\expandafter\ifx\csname l@#1\endcsname\relax
\typeout{** WARNING: IEEEtran.bst: No hyphenation pattern has been}%
\typeout{** loaded for the language `#1'. Using the pattern for}%
\typeout{** the default language instead.}%
\else
\language=\csname l@#1\endcsname
\fi
#2}}
\providecommand{\BIBdecl}{\relax}
\BIBdecl

\bibitem{Deniz2022}
D.~G{\"u}nd{\"u}z, Z.~Qin, I.~E. Aguerri, H.~S. Dhillon, Z.~Yang, A.~Yener,
  K.~K. Wong, and C.-B. Chae, ``Beyond transmitting bits: Context, semantics,
  and task-oriented communications,'' \emph{arXiv:2207.09353}, 2022.

\bibitem{semanticTheory}
Y.~Shao, Q.~Cao, and D.~Gunduz, ``A theory of semantic communication,''
  \emph{arXiv:2212.01485}, 2022.

\bibitem{JSCC2019}
E.~Bourtsoulatze, D.~B. Kurka, and D.~G{\"u}nd{\"u}z, ``Deep joint
  source-channel coding for wireless image transmission,'' \emph{IEEE Trans.
  Cognitive Commun. Netw.}, vol.~5, no.~3, pp. 567--579, 2019.

\bibitem{xie2021}
H.~Xie, Z.~Qin, G.~Y. Li, and B.-H. Juang, ``Deep learning enabled semantic
  communication systems,'' \emph{IEEE Trans. Signal Proc.}, vol.~69, pp.
  2663--2675, 2021.

\bibitem{Haotian}
H.~Wu, Y.~Shao, K.~Mikolajczyk, and D.~G{\"u}nd{\"u}z, ``Channel-adaptive
  wireless image transmission with \textsc{OFDM},'' \emph{arXiv:2205.02417},
  2022.

\bibitem{DeepJSCCq}
T.-Y. Tung, D.~B. Kurka, M.~Jankowski, and D.~G{\"u}nd{\"u}z,
  ``Deep\textsc{JSCC}-\textsc{Q}: Constellation constrained deep joint
  source-channel coding,'' \emph{arXiv:2206.08100}, 2022.

\bibitem{yang2022ofdm}
M.~Yang, C.~Bian, and H.-S. Kim, ``\textsc{OFDM}-guided deep joint source
  channel coding for wireless multipath fading channels,'' \emph{IEEE Trans.
  Cogn. Commun. Netw.}, 2022.

\bibitem{AttentionCode}
Y.~Shao, E.~Ozfatura, A.~Perotti, B.~Popovic, and D.~Gunduz, ``Attentioncode:
  Ultra-reliable feedback codes for short-packet communications,''
  \emph{arXiv:2205.14955}, 2022.

\bibitem{KO2021}
A.~V. Makkuva, X.~Liu, M.~V. Jamali, H.~Mahdavifar, S.~Oh, and P.~Viswanath,
  ``\textsc{KO} codes: Inventing nonlinear encoding and decoding for reliable
  wireless communication via deep-learning,'' in \emph{International Conference
  on Machine Learning}.\hskip 1em plus 0.5em minus 0.4em\relax PMLR, 2021, pp.
  7368--7378.

\bibitem{FLOAC}
Y.~Shao, D.~G\"und\"uz, and S.~C. Liew, ``Federated learning with misaligned
  over-the-air computation,'' \emph{IEEE Trans. Wireless Commun.}, vol.~21,
  no.~6, pp. 3951--3964, 2022.

\bibitem{PAPRsurvey}
Y.~Rahmatallah and S.~Mohan, ``Peak-to-average power ratio reduction in
  \textsc{OFDM} systems: A survey and taxonomy,'' \emph{IEEE communications
  surveys \& tutorials}, vol.~15, no.~4, pp. 1567--1592, 2013.

\bibitem{PAPRloss1}
M.~Kim, W.~Lee, and D.-H. Cho, ``A novel \textsc{PAPR} reduction scheme for
  \textsc{OFDM} system based on deep learning,'' \emph{IEEE Commun. Lett.},
  vol.~22, no.~3, pp. 510--513, 2017.

\bibitem{PAPRloss3}
L.~Li, C.~Tellambura, and X.~Tang, ``Improved tone reservation method based on
  deep learning for \textsc{PAPR} reduction in \textsc{OFDM} system,'' in
  \emph{IEEE Int. Conf. Wireless Commun. and Signal Proc.}, 2019.

\bibitem{PAPRloss4}
M.~Goutay, F.~A. Aoudia, J.~Hoydis, and J.-M. Gorce, ``Learning \textsc{OFDM}
  waveforms with \textsc{PAPR} and \textsc{ACLR} constraints,''
  \emph{arXiv:2110.10987}, 2021.

\bibitem{PAPRapprox}
R.~Van~Nee and A.~De~Wild, ``Reducing the peak-to-average power ratio of
  \textsc{OFDM},'' in \emph{IEEE Veh. Tech. Conf.}, vol.~3, 1998, pp.
  2072--2076.

\bibitem{SCFDMA}
H.~G. Myung, J.~Lim, and D.~J. Goodman, ``Single carrier \textsc{FDMA} for
  uplink wireless transmission,'' \emph{IEEE Veh. Tech. Maga.}, vol.~1, no.~3,
  pp. 30--38, 2006.

\bibitem{shaoIFDMA}
Y.~Shao and S.~C. Liew, ``Flexible subcarrier allocation for interleaved
  frequency division multiple access,'' \emph{IEEE Trans. Wireless Commun.},
  vol.~19, no.~11, pp. 7139--7152, 2020.

\bibitem{clip}
Y.-C. Wang and Z.-Q. Luo, ``Optimized iterative clipping and filtering for papr
  reduction of \textsc{OFDM} signals,'' \emph{IEEE Trans. Commun.}, vol.~59,
  no.~1, pp. 33--37, 2010.

\end{thebibliography}

\end{document}